\documentclass[conference]{IEEEtran}
\usepackage{blindtext, graphicx}
\usepackage[]{algorithm2e}
\usepackage{amsmath}
\ifCLASSINFOpdf
\else
\fi
\usepackage[bookmarks=true, bookmarksnumbered=true, bookmarksopen=true,
		unicode=true, colorlinks=true,
		linkcolor=blue,citecolor=blue,filecolor=blue,urlcolor=blue,
 		pdfstartview=FitH
		]{hyperref}

\usepackage[disable]{todonotes}
\newcommand{\todoVL}[1]{\todo[color=red!40, author=VL, inline]{#1}}
\newcommand{\todoRF}[1]{\todo[color=green!40, author=RF, inline]{#1}}

\begin{document}
%
\title{Relevance of Negative Links in Graph Partitioning: A Case Study Using Votes From the European Parliament}


\author{\IEEEauthorblockN{
	Israel Mendon\c{c}a, 
	Rosa Figueiredo,
	Vincent Labatut, 
	Philippe Michelon}
\IEEEauthorblockA{
	Universit\'e d'Avignon, LIA EA 4128, France\\
	Emails:\{mendonci, rosa.figueiredo, vincent.labatut, philippe.michelon\}@univ-avignon.fr}}

\maketitle

\begin{abstract}
In this paper, we want to study the informative value of negative links in signed complex networks. For this purpose, we extract and analyze a collection of signed networks representing voting sessions of the European Parliament (EP). We first process some data collected by the VoteWatch Europe Website for the whole $7^{th}$ term (2009-2014), by considering voting similarities between Members of the EP to define weighted signed links. We then apply a selection of community detection algorithms, designed to process only positive links, to these data. We also apply Parallel Iterative Local Search (Parallel ILS), an algorithm recently proposed to identify balanced partitions in signed networks. Our results show that, contrary to the conclusions of a previous study focusing on other data, the partitions detected by ignoring or considering the negative links are indeed remarkably different for these networks. The relevance of negative links for graph partitioning therefore is an open question which should be further explored.
\end{abstract}



%
\IEEEpeerreviewmaketitle

\section{Introduction}
\label{sec:introduction}
In \textit{signed} graphs, each link is labeled with a sign $+$ or~$-$, which indicates the nature of the relationship between the considered adjacent nodes. This type of graphs was primarily introduced in Psychology, with the objective of describing the relationship between people belonging to distinct social groups~\cite{heider}. More generally, a signed graph can be used to model any system containing two types of antithetical relationships, such as like/dislike, for/against, etc. This work and its extensions by Cartwright and Harary in the 1950's~\cite{dorwin-1,james-1,frank-1} are the basis for the concept of \textit{Structural balance}. A signed graph is considered \textit{structurally balanced} if it can be partitioned into two~\cite{dorwin-1} or more~\cite{james-2} mutually hostile subgroups each having internal solidarity. Here, the words \textit{hostile} and \textit{solidary} mean: \textit{connected by negative} and \textit{positive} links, respectively.

However, it is very rare for a real-world network to have a perfectly balanced structure: the question is then to quantify how balanced it is. For this purpose, one must first define a measure of balance, and then apply a method to evaluate the network balance according to this measure. For instance, one could consider counting the numbers of positive links located inside the groups, and of negative links located between them. Such a measure is expressed relatively to a graph partition, so processing the graph balance amounts to identifying the partition corresponding to the highest balance measure. In other words, calculating the graph balance can be formulated as an optimization problem.

By using different variants of the balance measure and/or by introducing some additional constraints, one can express various versions of the notion of balance. Each one potentially leads to a different optimization problem to be solved. However, besides the very classic measures such as the one mentioned previously, only a few recent works explored this aspects from an Operations Research perspective~\cite{SAC2015,rosa-t3,rosa-t5,bruggeman-1,brusco-2}. A deep investigation of efficient approaches and mathematical formulations to problems related with signed graph balance is therefore still missing. 


Independently from the Operations Research domain, the study and partition of signed graph has also recently been the object of several works in the domain of Complex Network Analysis, and more particularly community detection. The community detection problem originally concerns unsigned graphs. It consists in partitioning such a graph, in a way such that most links are located inside the groups (aka. communities) and only few remain between them. By definition, an unsigned graph focuses on a single type of relationships, say the positive ones. A signed graph representing the same system can therefore be considered as more informative, since it additionally contains the links of the other type (in our example, the negative ones). For this reason, a few authors tried to adapt existing community detection methods, in order to take advantage of this additional information~\cite{yang-1,leskovec-1,patrick-1,SAC2015}. 

Other authors tried to study how informative these additional links really are~\cite{paper-negtie}. Indeed, retrieving a signed network is a task potentially more costly than for an unsigned network, be it in terms of time, money, or methodological complications. For example, in the context of a ground survey, it is much easier to get people to name their friends than their foes. So, the question to know whether this extra cost is worth it is extremely relevant. In their work, Esmailian \textit{et al}.~\cite{paper-negtie} suggested that if one detects the communities based only on positive links (by ignoring negative ones), most negative links are already placed between the communities, and that the few ones located inside do not significantly affect the communities. The latter point is tested by checking that no additional division of the community allows increasing the overall balance. Consequently, using algorithms that do not take negative links into consideration, such as InfoMap~\cite{infomap-1}, it is possible to obtain a reasonably well partitioned network. However, we see two limitations to this work. First, in order to assess the significance of the negative links located inside the communities, Esmailian \textit{et al}. considered each community separately, instead of the graph as a whole. Second, these observations were made only for two datasets, both representing Social Networking Services (Slashdot and Epinions), so they do not necessarily apply to all networks, or event to all types of networks. 

In this paper, we want to explore further the informative value of negative links in the context of graph partitioning. To this purpose, we present a method to extract signed networks from voting data describing the activity of the \textit{Members of the European Parliament} (MEPs). Based on this new data, we apply state-of-the-art tools in order to partition the graph, on the one hand in terms of community structure, and on the other hand according to the notion of structural balance. We then compare the obtained partitions and show the presence of significant differences between them.

The contributions of this paper are essentially two-fold. First, we constitute a new dataset of signed networks and make it publicly available to the community, with the scripts used to obtain it. We treat the voting patterns using several parameters, leading to a collection of signed networks describing the behavior of MEPs according to various modes (time, topic...). Second, based on these data, we experimentally show that negative links \textit{can} be essential when partitioning networks. We see our work as complementary to that of Esmailian \textit{et al}., first because the use of a method taking negative links into account as a reference allows us to avoid the issue regarding the assessment of intra-community negative links; and second because we treat a different type of signed real-world networks, in which the links represent vote similarity instead of self-declared social relationships.

The rest of this paper is organized as follows. Section~\ref{sec:related_work} presents a review of the literature regarding the graph partition task. Section~\ref{sec:network_extraction} describes the method we used to extract signed networks from the raw data constituted of the sequences of MEPs votes. Section \ref{sec:partition_methods} summarizes the algorithms we selected to partition our signed networks. In Section~\ref{sec:experiments}, we present and discuss our experimental results regarding network extraction and network partition. Finally, we conclude by highlighting the main points of the article, and identifying some possible perspectives.


\section{Related Works}
\label{sec:related_work}
As mentioned before, the concepts of \textit{signed graph} and \textit{structural balance} were introduced by Heider~\cite{heider}. Later, Cartwright \textit{et al}.~\cite{dorwin-1} formalized Heider’s theory, stating that a balanced social group could be partitioned into two mutually hostile subgroups, each having internal solidarity. Observing that a social group may contain more than two hostile subgroups, Davis~\cite{james-2} proposed the notion of \textit{clusterable} signed graph. 


The clustering problem consists in finding the most balanced partition of a signed graph. Evaluating this balance according to the structural balance (SB) measure amounts to solving an optimization problem called \textit{Correlation Clustering} (CC) \cite{bansal-1}. This problem was addressed first by Doreian \& Mrvar~\cite{doreian-3}, who proposed an approximate solution and used it to analyze the structural balance of real-world social networks.
In~\cite{yang-1}, Yang \textit{et al}. called the CC problem \textit{Community Mining}, and proposed an agent-based heuristic called FEC to find an approximate solution. 
Elsner \& Schudy performed a comparison of several strategies for solving the CC problem in~\cite{elsner-1}, and applied them to document clustering and natural language processing issues. In this context, these authors identified the best strategy as a greedy algorithm able to quickly achieve good objective values with tight bounds. The solution of the CC problem and of some of its variants has already been used as a criterion to measure the balance of signed social networks~\cite{doreian-2,doreian-3,doreian-1,rosa-t5,SAC2015}, and as a tool to identify relations contributing to their imbalance~\cite{abell-1}. 
In~\cite{SAC2015}, the authors provide an efficient solution of the CC problem, by the use of a ILS metaheuristic. The proposed algorithm outperforms other methods from the literature on 3 huge signed social networks. In this work, we will use this tool to evaluate the imbalance of the MEPs networks.


In the complex networks field, works dedicated to signed networks focus only on the clustering problem, as defined by Davis~\cite{james-2}. 
\todoVL{@all: I'm actually not so sure of that anymore, I must check again. Community detection involves taking link density into account, which is ont considered in the CC problem (cf. the article of Esmailian \textit{et al}.}
Various methods were proposed for this purpose: evolutionary approaches~\cite{yadong-1,liu-1,yang-2,yujie-1}, agent-based systems~\cite{yang-1}, matrix transformation~\cite{yang-3}, extensions of the Modularity measure~\cite{amelio-1,bruggeman-1,gomez-1,macon-1,traag-1}, simulated annealing~\cite{bogdanov-1}, spectral approaches~\cite{pranay-1,kunegis-1,wu-1}, particle swarm optimization~\cite{cai-1,gong-1}, and others. Some authors performed the same task on bipartite networks~\cite{mvrar-1}, while others relaxed the clustering problem in order to identify overlapping communities~\cite{chen-1}. Although the methods listed here were applied to networks representing very different systems, authors did not investigate the possibility that some alternative versions of the clustering problems were more appropriate to certain data.

Few works tried to compare the CC and community detection approaches. As mentioned in the introduction, Esmailian \textit{et al} showed that, in certain cases, partitions estimated in signed networks by community detection methods, i.e. based only on the positive links, can be highly balanced \cite{paper-negtie}. However, this work was conducted only on two networks of self-declared social interaction networks (Epinions and Slashdot), and using a single community detection method (InfoMap \cite{infomap-1}). Moreover, they did not compare their results to partitions detected by algorithms designed to solve the CC problem. We investigate if this statement also holds for other real-world networks and community detection methods, and how these compare to results obtained with CC methods.

\section{Network Extraction}
\label{sec:network_extraction}
In this section, we describe the source we used to retrieve our raw data, and the process we applied to extract signed networks from these data.

\subsection{European Parliament Votes}
\label{sec:datasets}
In order to be able to conduct our experiments, we were looking for data allowing to extract some form of signed network of interactions. Moreover, in future works, we want to study how the network and the structural balance evolve, so the data had to be longitudinal, with stable nodes (nodes should not change too much through time). The best data we could find relatively to these criteria are those describing the activity of the \textit{European Parliament}\footnote{\url{http://www.europarl.europa.eu/}}. More precisely, we focused on the votes of the Members of the European Parliament (MEPs).

The Website \textit{VoteWatch Europe}\footnote{\url{http://www.votewatch.eu/}} is a non-partisan international non-governmental organization, completely independent from national and local governments, from the European Union, as well as from political parties, institutions, agencies, businesses and all other bodies. Their goal is to provide easy access to the votes and other activities of the European parliament (among other European institutions). 
Votewatch compiles data provided by the EP to give a full overview of the MEPs activity. In particular, they describe the vote cast by each MEP for each document considered at the EP. Each MEP is also described through his name, country and political group. Other fields are available too, which we have not used yet, such as how loyal the MEP is to his political group. To summarize, the behavior of a MEP is represented by the series of votes he cast over a certain time period (e.g. a year, a term).

\todoVL{@IM: you didn't really separate data from processing. a lot of things regarding the data were in subsection B, when they are actually controlled by votewatch and therefore do not depend on you and your process (rather the opposite). so I moved them here.}

For a given document, a MEP can express his vote in one of the three following ways: \textsc{For} (the MEP wants the document to be accepted), \textsc{Against} (he wants the document to be rejected) and \textsc{Abstain} (he wants to express his neutrality). Besides these expressed votes, it is also possible for the MEP not to vote at all, leading to the following possibilities: \textsc{Absent} (the MEP was not present during the vote), \textsc{Did not Vote} (he was there, but did not cast his vote), and \textsc{Documented Absence} (he was not there but justified his absence).

\begin{table}[h]
	\centering
	\caption{List of all policies relative to the documents voted at the European Parliament, with the corresponding numbers of documents, for the 7th term}
	\label{tab:policies}
	\begin{tabular}{|l|r|}
		\hline
		\textbf{Policy} & \textbf{Number of documents} \\
		\hline
	    Agriculture & 53\\
	    \hline
	    Budget & 179\\
	    \hline
	    Budgetary control & 113\\
	    \hline
	    Civil liberties, justice \& home affairs & 99\\
	    \hline
	    Constitutional and inter-institutional affairs & 40\\
	    \hline
	    Culture \& education & 19\\
	    \hline
	    Development & 29\\
	    \hline
	    Economic \& monetary affairs & 128\\
	    \hline
	    Employment \& social affairs & 44\\
	    \hline
	    Environment \& public health & 100\\
	    \hline
	    Fisheries & 53\\
	    \hline
	    Foreign \& security policy & 191\\
	    \hline
	    Gender equality & 28\\
	    \hline
	    Industry, research \& energy & 51\\
	    \hline
	    Internal market \& consumer protection & 39\\
	    \hline
	    Internal regulations of the EP & 7\\
	    \hline
	    International trade & 106\\
	    \hline
	    Legal affairs & 67\\
	    \hline
	    Petitions & 5\\
	    \hline
	    Regional development & 35\\
	    \hline
	    Transport \& tourism & 40\\
	 	\hline
	\end{tabular}
\end{table}

For each document, we also have access to the category it belongs to, called \textit{Policy}. It corresponds roughly to the main theme treated in the considered document. All the policies treated during the 7th term of the EP are listed in Table \ref{tab:policies}, with the numbers of documents they concern. 

VoteWatch gives us access to raw data, which could be described as individual data, in the sense they describe the state and behavior of the MEPs when considered independently from each others. However, a network is by nature relational, i.e. it represents the relationships between some objects of interest. Thus, we need to process the VoteWatch data in order to retrieve the networks we want.

\subsection{Extraction Process}
\label{sec:agreement}
Our extraction process is 
two-stepped. As mentioned before, in the data received from VoteWatch, the behavior of each MEP is represented by a series of votes, corresponding to all the documents reviewed by the EP during one term. In this article, we focused on the 7th term (from june 2009 to june 2014). We first filter these data depending on temporal and topical criteria. In other terms, if required, it is possible to focus only on the documents related to a specific policy and/or a specific period of the term, for instance a given year. The second step consists in comparing individually all MEPs in terms of similarity of their voting behaviors. The result of this process is what we call the \textit{agreement} matrix $M$. Each numerical value $m_{uv}$ contained in the matrix represents the average agreement between two MEPs $u$ and $v$, i.e. how similarly they vote over all considered documents. 

The filtering step is straightforward, however the agreement processing constitutes a major methodological point: depending on how it is conducted, it can strongly affect the resulting network. For a pair of MEPs $u$ and $v$ and a given document $d_i$, we define the \textit{document-wise agreement score} $m_{uv}(d_i)$ by comparing the votes of both considered MEPs. It ranges from $-1$ if the MEPs fully disagree, i.e. one voted \textsc{For} and the other \textsc{Against}, to $+1$ if they fully agree, i.e. they both voted \textsc{For} or \textsc{Against}. 

However, as we mentioned previously, a vote can take other values than just \textsc{For} and \textsc{Against}, and those must also be treated. Let us consider first the non-expressed votes: \textsc{Absent}, \textsc{Did not Vote} and \textsc{Documented Absence}. The EU distinguishes these different forms of absence not for political, but rather for administrative reasons, so we decided to consider them all simply as absences. The common approach when treating this type of vote data \cite{Porter2005,Maso2014} is to ignore all documents $d_i$ for which at least one of the considered MEPs was absent. However, certain MEPs are absent very often, which mean they would share a very small number of documents with others. This approach could therefore artificially produce extremely strong agreement or disagreement scores. To avoid this, we assign a neutral score of $0$ when at least one person is absent for a given document.

Handling the abstentions is a bit trickier, because such a behavior can mean different things. A MEP can choose not to vote because he is personally \textsc{For} or \textsc{Against}, but undergoes some pressure (from his political group, his constituents, etc.) to vote the other way: in this case, voting \textsc{Abstain} is a way of expressing this conflicting situation. But abstaining could also simply represent neutrality, meaning the MEP is neither \textsc{For} nor \textsc{Against} the considered document. There is no consensus in the literature, and different approaches were proposed to account for \textsc{Abstain}-\textsc{For}, \textsc{Abstain}-\textsc{Against} and \textsc{Abstain}-\textsc{Abstain} situations \cite{macon-1,Porter2005,Maso2014}. Here, we present two variants corresponding to different interpretations. In the first, described in Table \ref{tab:tabela_1}, an abstention is considered as half an agreement with \textsc{For} or \textsc{Against}, leading to a score of $+0.5$. In the second, described in Table \ref{tab:tabela_2}, two abstaining MEPs are considered to fully agree ($+1$). But, when only one of them abstains, we consider there is not enough information to determine whether they agree or disagree, and we therefore use a $0$ score. Note absences were left out of the tables for clarity.


\begin{table}[h]
	\centering
	\caption{Vote weights representing abstention as half an agreement}
	\label{tab:tabela_1}
	\begin{tabular}{|l|c|c|c|}
		\hline
	 	 & \textsc{For} & \textsc{Abstain} & \textsc{Against} \\
	 	\hline
	 	\textsc{For}     & $+1$   & $+0.5$ & $-1$   \\
	 	\textsc{Abstain} & $+0.5$ & $+0.5$ & $+0.5$ \\
	 	\textsc{Against} & $-1$   & $+0.5$ & $+1$   \\
	 	\hline
	\end{tabular}
\end{table}

\begin{table}[h]
	\centering
	\caption{Vote weights representing abstention as an absence of opinion}
	\label{tab:tabela_2}
	\begin{tabular}{|l|c|c|c|}
		\hline
 		 & \textsc{For} & \textsc{Abstain} & \textsc{Against} \\
 		\hline
 		\textsc{For}     & $+1$ & $0$  & $-1$ \\
 		\textsc{Abstain} & $0$  & $+1$ & $0$  \\
 		\textsc{Against} & $-1$ & $0$  & $+1$ \\
 		\hline
	\end{tabular}
\end{table}

The document-wise agreement score is completely defined by selecting one of the proposed tables. The average agreement is then obtained by averaging this score over all considered documents. More formally, let us consider two users $u$ and $v$ and note $d_1,...,d_\ell$ the documents remaining after the filtering step, and for which $u$ and $v$ both cast their votes. The \textit{average agreement} $m_{uv}$ between these two MEPs is:
\begin{equation}
	 m_{uv}=\frac{1}{\ell} \sum_{i=1}^{\ell} m_{uv}(d_i)
\end{equation}

\section{Partition Methods}
\label{sec:partition_methods}
In this section, we present the methods used to partition the signed network extracted from the VoteWatch data. We first introduce the community detection approaches we selected for our experiment. Then we formally define the Correlation Clustering problem and describe the algorithm we used in this article to estimate its solutions.

\subsection{Community Detection}
\label{sec:communities}
In the literature, the problem of community detection is usually defined in an informal way. It consists in finding a partition of the node set of a graph, such that many links lie inside the parts (communities) and few lie in-between them. An other way of putting it is that we are looking for groups of densely interconnected nodes, relatively to the rest of the network \cite{santo-1}. It is difficult to find a formal definition of this problem, or rather, to find a \textit{unique} formal definition: many authors present and solve their own variant. Because of this, the algorithms presented in the literature do not necessarily solve the exact same problem, although it is still named community detection. To account for this variance, we selected several methods for our experiments, which we briefly present here. All of them are able to process weighted networks.

\textbf{InfoMap \cite{infomap-1}.} The community structure is represented through a two-level nomenclature based on Huffman coding: one level to distinguish communities in the network and the other to distinguish nodes in a community. The problem of finding the best community structure is expressed as minimizing the quantity of information needed to represent some random walk in the network using this nomenclature. With a partition containing few intercommunity links, the walker will probably stay longer inside communities; therefore only the second level will be needed to describe its path, leading to a compact representation. The authors optimize their criterion using simulated annealing.


\textbf{EdgeBetweenness~\cite{newman-2}.} This divisive hierarchical algorithm adopts a top-down approach to recursively split communities into smaller and smaller node groups. The split is performed by iteratively removing the most central link of the network. This centrality is expressed in terms of edge-betweenness, i.e. number of shortest paths running through the considered link. The idea behind this method is that links connecting different communities tend to be present in the many shortest paths connecting one community to the other. Once the network has been split in two separate components, each one is split again applying the same process, and so on. The resulting components correspond to communities in the original network.

\textbf{WalkTrap}~\cite{pascal-1}. To the contrary of EdgeBetweenness, this is an agglomerative hierarchical algorithm, which means it uses a bottom-up approach to merge communities into larger and larger groups, starting from singletons. To select the communities to merge, WalkTrap uses a random walk-based distance. Indeed, random walkers tend to get trapped into communities, because most locally available links lead to nodes from the same communities, while only a few links all to escape this community (by definition). If two nodes $u$ and $v$ are in the same community, the probability to reach, through a random walk, a third node located in the same community should not be very different for $u$ and $v$. The distance is constructed by summing these differences over all nodes, with a correction for degree.

\textbf{FastGreedy~\cite{clauset-1}.} Like WalkTrap, this algorithm adopts an agglomerative hierarchical approach. But this time, the merges are not decided using a distance measure, but rather by locally optimizing the well-known objective function called \textit{Modularity} \cite{Newman2006}. Briefly, this measure compares the proportion of intra-community links present in the network of interest, to the expectation of the same quantity for a randomly generated network of similar size and degree distribution. The process stops when it is not possible to improve the modularity anymore, or when there is no more communities to merge.

\subsection{Correlation Clustering}
\label{sec:ccp}
Before formally describing the CC problem, we need to introduce some notations and definitions first. Let $G=(V,E,s,w)$ be a weighted undirected signed graph. The sets $V$ and $E$ correspond to the nodes and links constituting the graph. The functions $s:E\rightarrow\{+,-\}$ and $w:E\rightarrow[0;+1]$ assign a sign and a positive weight to each link in $E$, respectively.

\todoVL{@IM+RF: in this part, I removed some parts which did not seem necessary. I fixed some notations, and I rewrote the numbered equations, I think the negative weight sum was incorrect. Rosa, please review my modifications, maybe I introduced some errors.}

A link $e\in E$ is called {\it negative} if $s(e)=-$ and {\it positive} if $s(e)=+$. Let $E^- \subset E$ and $E^+ \subset E$ denote the sets of negative and positive links in $G$, respectively. Notice that, according to the above definitions, $E=E^-\cup E^+$. We define the negative and positive subgraphs of $G$ as $G^-=(V,E^-)$ and $G^+=(V,E^+)$, respectively. The \textit{complementary negative graph} is $\overline{G^-} = (V,\overline{E^-})$, where $\overline{E^-} = \mathcal{P}_2(V) \setminus E^-$, $\mathcal{P}_2(V)$ being the set of all unordered pairs from $V$.



Let us consider a partition $P$ of $V$ such that $P = \{V_1,...,V_k\}$. A link is said to be \textit{cut} if it connects nodes from two different parts. We note $E[V_i:V_j] \subset E$ the set of links connecting two nodes from $V_i$ and $V_j$ (cut links), and $E[V_i] \subset E$ the set of links connecting two nodes from $V_i$ (so, $E[V_i] = E[V_i:V_i]$) (uncut links). 

As mentioned before, negative links located inside parts (uncut negative links) and positive links located between parts (cut positive links) are considered to lower the graph balance. For $V_i$, the total weight of \textit{uncut negative links} is:
\begin{equation}
	\displaystyle \Omega^{-}(V_i) = \sum_{e \in E^{-}\cap E[V_i]} w_e
\end{equation}

\noindent And for two parts $V_i$ and $V_j$, the total weight of \textit{cut positive links} $\Omega^{+}$ is:
\begin{equation}
	\displaystyle \Omega^{+}(V_i,V_j) =  \sum_{e \in E^{+}\cap E[V_i:V_j]} w_e
\end{equation}

%
%

\noindent The \textit{Imbalance} $I(P)$ of a partition $P$ can be defined as the sum of uncut negative and cut positive links over the whole graph:

\begin{equation}
	\label{eq:Imbalance}
	\displaystyle I(P) = \sum_{1 \leq i \leq k} \Omega^{-}(V_i) + \sum_{1 \leq i < j \leq k} \Omega^{+}(V_i,V_j)
\end{equation}


Finally, the \textit{Correlation Clustering problem} is the problem of finding a partition $P$ of $V$ such that the \textit{imbalance} $I(P)$ is minimized.


In this work we will solve the CC problem using the \textit{Parallel ILS algorithm} presented in~\cite{SAC2015}, which was designed to solve the CC problem in large real-world networks. ILS is itself a metaheuristic approach allowing to obtain good quality solutions by applying iteratively greedy search methods \cite{Lourenco2010}. Starting from an initial solution estimated through a greedy method, the general principle is two-stepped: first, some perturbations are introduced to modify the current best solution; second, some local searches are performed to find better solutions within the neighborhood. This iterative process is stopped when some condition is met (minimal quality, time limit, etc.). This specific implementation is parallelized, in order to improve speed.


Considering that the networks extracted from the VoteWatch data are very dense, we had to perform some minor modifications on the original Parallel ILS algorithm, so that the processing time was acceptable. First, the search space used in the local search was reduced by adding a probably ($0.7$) of visiting a neighbor solution. In other terms, in average we limit the search to only a part of the neighborhood. Second, the perturbation level had to be reduced to $15$, half of the maximum run number in the original work.
\todoVL{@IM+RF: What is the unit of the perturbation level ? $15$ what ?}
\todoVL{@all: also, we need to explain that we tested different values of these parameters.}





\section{Results and Discussion}
\label{sec:experiments}
In this section, we first describe the networks extracted from the VoteWatch data, and how they are affected by the parameters controlling this extraction process. Then, we discuss the results obtained with the partition methods presented in section \ref{sec:partition_methods}.

In order to process the VoteWatch data, we developed a tool called \textit{NetVotes}, which takes the form of a collection of R scripts. It implements the method described in section \ref{sec:agreement}, and additionally calculates some metrics describing the studied networks and their partitions. It is generic enough to treat any type of data of the same form. To perform the community detection, we used the igraph R package, which contains all the algorithms we selected. For the CC problem, we used the author's version of Parallel ILS, which we modified as explained in section \ref{sec:ccp}. All our source code, as well as the data it outputs, are publicly available on GitHub\footnote{\url{https://github.com/CompNet/NetVotes/}} and FigShare\footnote{\url{http://figshare.com/articles/NetVotes_Data/1456268}}, respectively.



\subsection{Networks Extraction}
\label{subsec:networks_creation}
As described in section \ref{sec:agreement}, our extraction method takes three parameters: the table used to process the agreement scores, the policy and the time period. We proposed $2$ different tables, there are $21$ policies and we also considered all documents independently from their policies, and we considered each year separately as well as the whole $5$-year long $7^th$ term (2009-2014). This amounts to a total of $264$ different modalities. However, in certain cases, the filtering step led to less than $2$ documents, so we were not able to extract networks for all combinations of policies and time periods.

\todoVL{@IM: about this 264 $\rightarrow$ you had put 132 networks, but I suppose you didn't count the fact there are two different tables? or am I missing something? maybe is it related to some policies+periods leading to less than 2 documents? (last sentence above)}

\begin{figure}[!ht]
    \center
    \includegraphics[width=3.5in]{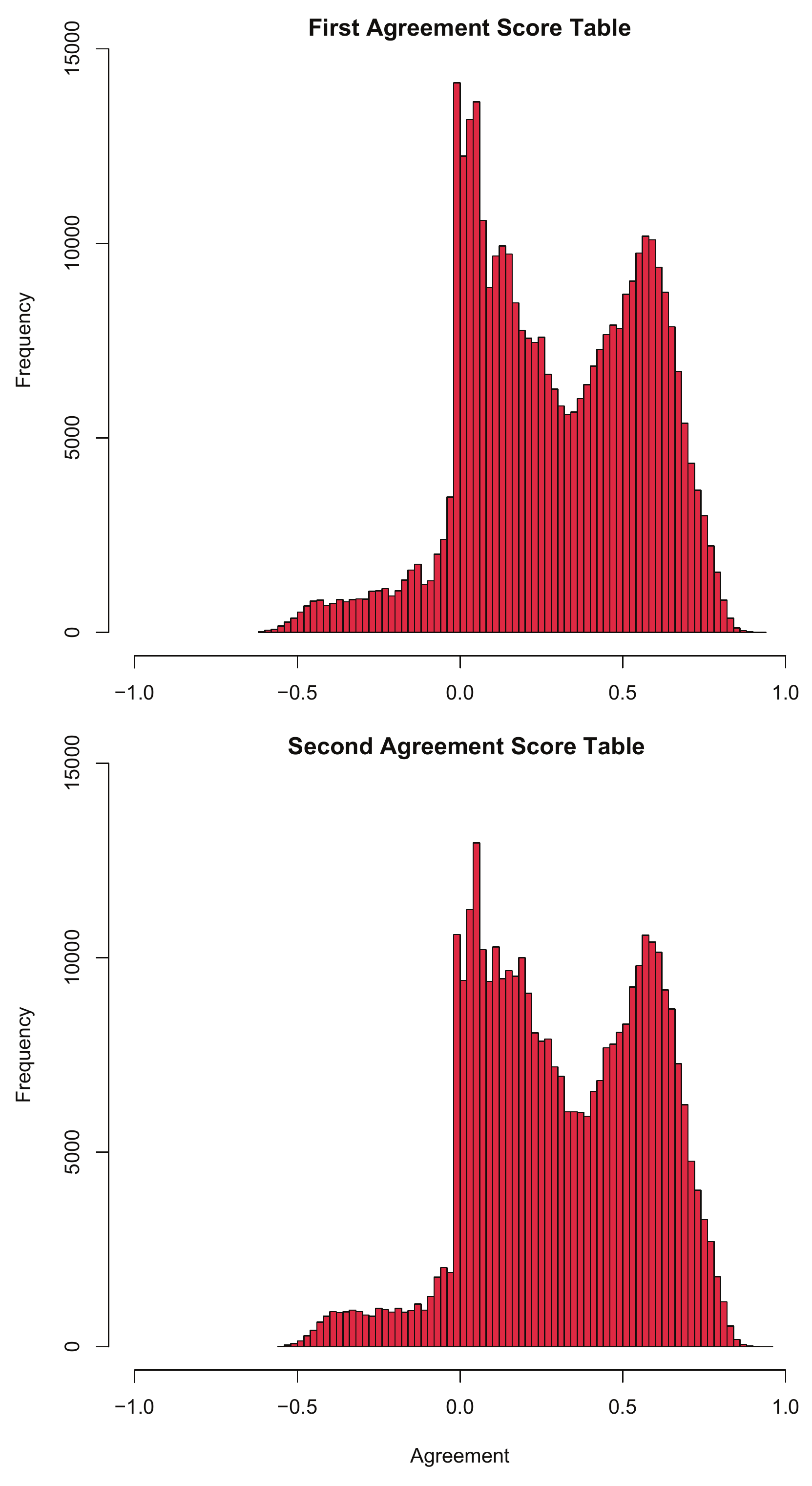}
    \caption{Agreement distribution for the whole term and all policies when using Table \ref{tab:tabela_1} (top) and Table \ref{tab:tabela_2} (bottom)}
    \label{fig:AgreementDitribution}
\end{figure}

We first study how the choice of the table used to process the agreement scores affects the extracted network. Figure~\ref{fig:AgreementDitribution} shows the average agreement distribution for Table \ref{tab:tabela_1} (top plot) and Table \ref{tab:tabela_2} (bottom), using all documents for the whole term (i.e., not applying any filter). Both distributions are very similar, with a clear separation between the negative and positive values. The agreement side is bimodal, with larger frequencies around $0$ and $0.6$--$0.7$. The right peak can be explained by the fact the majority of MEPs tend to vote similarly most of the time. The other peak, located at zero, is due to the frequent absence of a certain number of MEPs. When a MEP is absent for a given document, his agreement score with all other MEPs is zero. It is not as contrasted on the disagreement side, with a much flatter distribution. Moreover, there is no strong disagreement since the smaller values are around $-0.5$ (by comparison, the agreement values can get close to $0.9$). This means only a small proportion of MEPs systematically disagree with the rest of the EP. 

The same observations can be made when considering the different policies independently, as well as when considering each year separately. There are some variations in terms of position and amplitude of the right peak, but this is mainly due to large differences in the number of documents discussed for each policy. The nature of the table used to process the agreement scores does not seem to have any clear effect on the average agreement distribution. We additionally tried to use a variant of Table \ref{tab:tabela_1}, replacing the $+0.5$ (half-agreement) by $-0.5$ (half-disagreement) for the situations involving \textsc{Abstain} vs. \textsc{For} or \textsc{Against}. The results where extremely similar, confirming our observation. Consequently, in the rest of the article, we present only the results obtained with Table \ref{tab:tabela_2}.

\subsection{Partition Comparison}
\label{sec:comparison_algorithms}
We now want to study how the selected community detection methods behave on the CC problem, when compared to Parallel ILS, an algorithm specifically designed to treat this problem. However, as mentioned in Section \ref{sec:communities}, these methods can only take positive links into account, so they cannot be applied directly to our signed networks, unlike Parallel ILS.
To solve this issue, we proposed to consider two subgraphs of the original signed networks: the signed graph and the complementary negative graph, noted $G^+$ and $\overline{G^-}$ in Section \ref{sec:ccp}, respectively. The former is a version of the original graph retaining only its positive links. The latter contains all possible links but the ones labeled negative in the original graph. In both cases, the result is a graph with only one type of unlabeled links, representing a part of the information originally conveyed by the original graph. This is very consistent with our objective, since we want to study if the information loss translates in terms of detected partitions.

\todoVL{@IM: some stuff (notations, and son on) here was already in the CC section, so I removed it. The part relative to the number of nodes should not be here, but rather in the previous subsection where you present the graphs, so I moved it.}

We applied all the selected community detection algorithms to both types of graphs, for all the modalities described in the previous subsection. For space matters, it is not possible to display and comment all of them, so we decided to focus on two policies of interest, over all years and over the whole term. We picked \textit{Foreign \& Security Affairs} because it is the most frequent, with 191 documents, and \textit{Agriculture} because it is also well represented (51 document) and topically very distant from the former. 
\todoVL{@IM: I completely disagree with agriculture being non-conflictual. There were huge tensions between France and UK/germany because of the way the UE gives money to farmers. So I adapted your argument here.}

\begin{figure}[!ht]
    \center
    \includegraphics[width=3.5in]{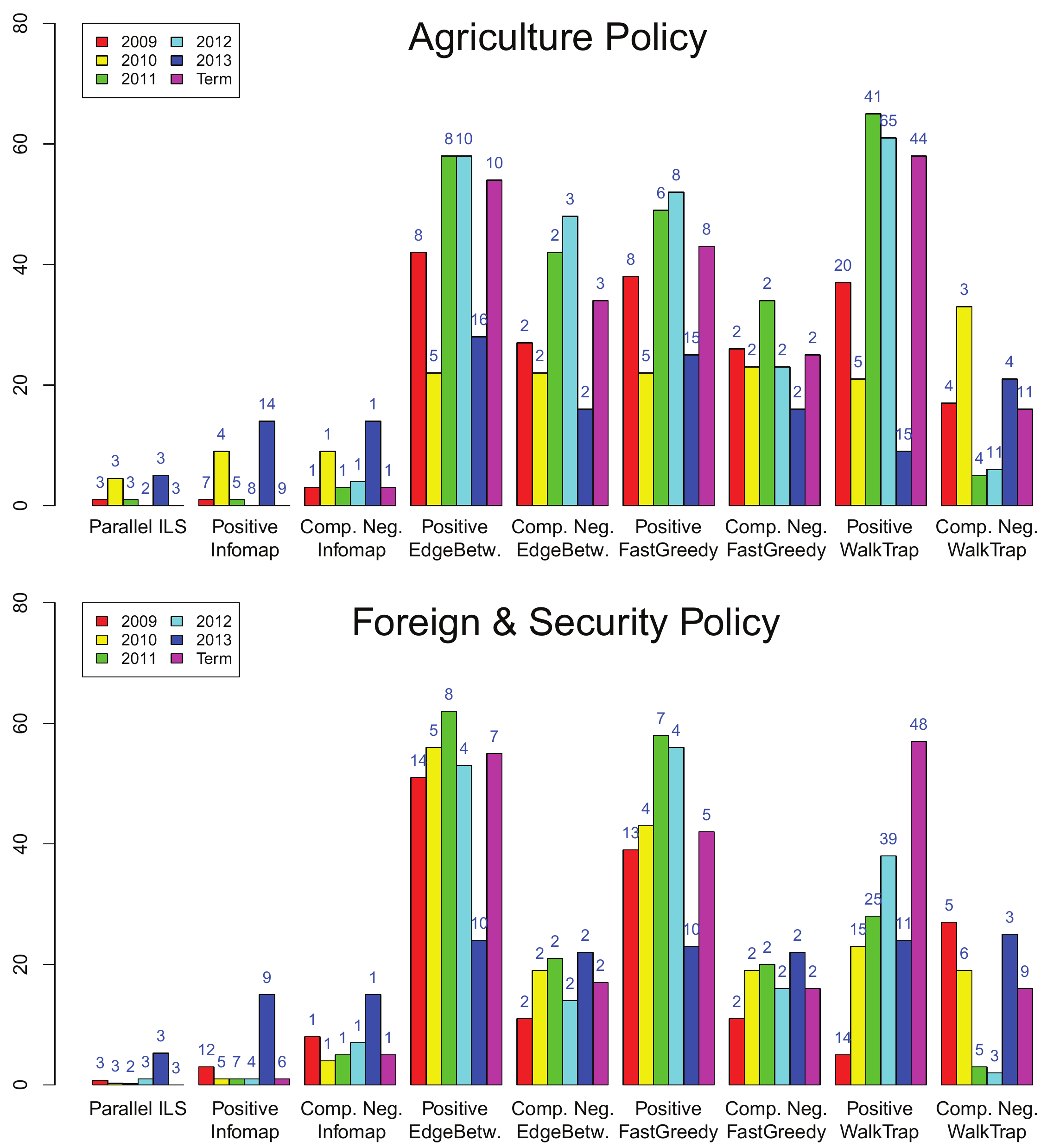}
    \caption{Imbalance of the partitions (bars) and numbers of detected clusters (blue values), obtained through Parallel ILS (left bar group) and community detection methods (other bar groups), for each year and the whole term (see legend), processed for the \textit{Agriculture} (top plot) and \textit{Foreign \& Security} (bottom plot) policies
   \todoVL{@IM: in this plot the "multilevel" algo was mentioned (i.e. Louvain algorithm, its real name), but the EdgeBetweenness was not mentioned. This was not consistent with the description in section "community detection", so I corrected accordingly.}
	}
    \label{fig:Imbalance}
\end{figure}

The obtained results are shown in Figure \ref{fig:Imbalance}. The top plot is dedicated Agriculture and the bottom one to Foreign \& Security Affairs. Each group of bars represents the results obtained by one algorithm for each year taken independently, and for the whole term (see the legend). The bar heights are proportional to the imbalance of the estimated partitions, as described in equation (\ref{eq:Imbalance}), only they are expressed in terms of percents relatively to $|E|$. The numbers on top of the bars indicate how many parts (communities) the corresponding partitions contain. Note the displayed results are representative of the other policies.

Let us compare the algorithms performances. EdgeBetweenness, FastGreedy and WalkTrap are far from finding optimal results when processing the positive subgraphs: they obtain scores ranging from $20\%$ to more than $60\%$ imbalance, and generally find a high number of clusters. The multitude of clusters is certainly the cause for these large imbalances. Note this observation is not inconsistent with being efficient at detecting communities, since this task implies taking link density into account. The behavior of the same algorithms is very different when applied to the complementary negative subgraph. The number of detected clusters is much smaller (generally around $2$--$5$), and the imbalance is smaller, but still around $20\%$. The reason for that is certainly that the graphs being much denser, it becomes harder to distinguish dense subroups, i.e. communities.

The InfoMap algorithm is much more successful at detecting balanced partitions, and reaches much smaller imbalance than the other community detection algorithms (always less than $20\%$, often less than $5\%$). However, on the negative complementary graphs, InfoMap simply puts all the nodes in the same cluster, so these results cannot be considered as relevant. On the positive graphs, the imbalance is very low (with the exception of the year 2013), close to $1\%$, and the algorithm finds $4$--$14$ clusters. The results obtained with Parallel ILS are even better, in terms of imbalance, since they consistently get close to $0\%$. Moreover, the number of clusters is relatively low ($2$--$3$), which corresponds to what we were expecting \textit{a priori}. Indeed, the EP is known to be split in two major political sides (EPP and S\&{}P), with some punctual alliances of smaller parties, leading to the formation of third or fourth groups. It is worth noticing that the imbalance is more marked for both algorithms for the year 2013, for both considered policy. This might be due to this year being the last in the $7^{th}$ term, and therefore coinciding with the negotiation of the $8^{th}$ term budgets and changes in the policies orientation. For instance, the CAP (Common Agricultural Policy) was made greener\footnote{\url{http://www.europarl.europa.eu/pdfs/news/expert/infopress/20131118IPR25538/20131118IPR25538_en.pdf}}. Such changes lead to stronger discussions in the EP, and may challenge the balance of certain political groups. 

In average, InfoMap identifies partitions 3 times more imbalanced than Parallel ILS and also tends to partition the graph in more clusters. Table \ref{tab:NMI} compares the InfoMap and Parallel ILS partitions in terms of \textit{Normalized Mutual Information}, which is the standard measure to compare partitions in the domain of unsupervised classification \cite{Fred2003}. This measure ranges from $-1$ (completely different) to $+1$ (completely identical), whereas $0$ represents statistical independence. The values obtained for both considered policies, and for all the time periods, are extremely close to zero. This means the partitions detected by the two algorithms have little in common, even though their number of clusters and/or imbalance level are sometimes similar.

\begin{table}[h]
	\centering
	\caption{Comparison of the InfoMap and Parallel ILS partitions in terms of NMI}
	\label{tab:NMI}
	\begin{tabular}{|l|r|r|r|r|r|r|}
		\hline
		\textbf{Policy} & \textbf{2009} & \textbf{2010} & \textbf{2011} & \textbf{2012} & \textbf{2013} & \textbf{Term} \\
	 	\hline
	 	Agriculture & $0.01$ & $0.04$ & $0.01$ & $0.02$ & $0.01$ & $0.02$ \\
	 	\hline
	 	Foreign Affairs & $0.01$ & $0.01$ & $0.01$ & $0.01$ & $0.03$ & $0.02$ \\
	 	\hline
	\end{tabular}
\end{table}

We can conclude by stating that, on these data, our results do not confirm the findings of Esmailian \textit{et al}. \cite{paper-negtie} regarding the low informative value of negative links. Taking negative links into account leads to a lower imbalance and a different partition, containing larger clusters. Moreover, among our selection of community detection algorithms, InfoMap is the only one to exhibit a behavior comparable to that of Parallel ILS. This means the notion of community implemented in this algorithm, which relies on an information compression-based approach, can be considered as compatible enough with the concept of structural balance. However, this is not the case for the other considered methods, based on link centrality, node distance and modularity. Discussing collectively the different methods proposed to solve the community detection problem might not be relevant, since the notions of community they rely upon are different (despite a common name).


\todoRF{@all It would be good to close this section with an analyse of our (better) results (ils algorithm) on the votes on the EP. For ex: we have something to say about the balance of the network on different years? We have more imbalance in the last year for both subjects? We could show more results of mario alg and infomap on other subjects?}
\todoVL{@all: agreed, I tried to add some example of what we could do, above the NMI discussion}

\section{Conclusion}
\label{sec:conclusion}
In this article, we have investigated some of the aspects inherent to the partition of signed networks, using data from the European Parliament (EP). We first extracted a collection of networks using the voting patterns of the Members of the EP. Then, we applied a selection of community detection methods to these networks, as well as Parallel ILS, an algorithm specifically designed to treat signed graphs. Among the former, the best results in terms of structural balance are obtained, by far, by InfoMap. However, in average, Parallel ILS detected partitions three times more balanced. This seems to be due to the fact community detection methods ignore negative links and focus instead on link density. Independently from the balance aspect, the number of clusters detected by ILS is lower, which is more consistent with the studied system. 

These results are in opposition with the finding of Esmailian \textit{et al}.~\cite{paper-negtie}, however they do not invalidate them. Indeed, in both cases, the experiments were performed on a very limited number of networks. The process should be conducted on a large number of different datasets in order to draw more reliable conclusions. In our future work, we plan to constitute a collection of real-world signed networks in order to perform this task. We also want to continue studying the MEPs voting data in further details, focusing on the interpretation of the identified balanced clusters.

\todoVL{@all: should not forget the acknowledgments (mario + lab money)}

\ifCLASSOPTIONcaptionsoff
  \newpage
\fi



%

%

\bibliographystyle{unsrt} 
\bibliography{bibliography} 




\end{document}